\documentclass[twocolumn]{aastex631}

\usepackage[hyphens]{url}    

\shorttitle{ixpe deadtime}
\shortauthors{Vivekanand}

\begin{document}

\title{Phase resolved deadtime of the Crab pulsar using IXPE data}

\author{M. Vivekanand}
\affiliation{No. $24$, NTI Layout $1$\textsuperscript{st} Stage, $3$\textsuperscript{rd} Main,
$1$\textsuperscript{st} Cross, Nagasettyhalli, Bangalore $560094$, India.}

\email{viv.maddali@gmail.com}

\begin{abstract}

After receiving an X-ray photon, an X-ray detector is not operational for a duration known
as deadtime. It is detector specific and its effect on the data depends upon the luminosity 
of the source. It reduces the observed photon count rate in comparison to the expected one.
In periodic sources such as the Crab pulsar, it can distort the folded light curve (FLC).
An undistorted FLC of the Crab pulsar is required in combination with its polarization 
properties for studying its X-ray emission mechanism. This work derives a simple formula 
for the distortion of the FLC of a pulsar caused by the detector deadtime, and validates 
it using Crab pulsar data from the X-ray observatories {{\it NICER}} and {{\it NUSTAR}}, 
which have very small and relatively large detector deadtimes respectively. Then it 
derives a method for correcting the distorted FLC of the Crab pulsar in {{\it IXPE}} data, 
which has intermediate detector deadtime. The formula is verified after addressing several 
technical issues. This work ends with a discussion of why an undistorted FLC is important 
for studying the formation of cusps in the FLC of the Crab pulsar.

\end{abstract}

\keywords{Stars: neutron -- Stars: pulsars: general -- Stars: pulsars: individual PSR J0534+2200 -- Stars: pulsars: individual PSR B0531+21 -- X-rays: general}

\section{Introduction} \label{sec:intro}

X-ray detectors have a duration known as the detector deadtime which occurs after reception of 
an X-ray photon, during which they are inoperative. If the detector deadtime is much smaller
than the mean time interval between X-ray photons from the source, which is the inverse of its 
expected count rate, then the observed count rate is also the same. As this mean time interval
decreases in comparison to the detector deadtime, i.e. for more luminous sources, the ratio of 
the observed to the expected count rate decreases. The observed count rate is defined as the 
ratio of the observed number of X-ray photons $N_o$ and the duration $\Delta T$ in which they 
are observed. Now $\Delta T$ is the sum of the durations $\Delta T_d$ and $\Delta T_l$, which 
are the cumulative sub-durations in $\Delta T$ in which the detector is dead and live, 
respectively. The expected count rate is the ratio $N_o / \Delta T_l$, since $\Delta T_l$
is the actual time spent observing the source. This is larger than the observed count rate 
$N_o / \Delta T$.

For periodic sources such as the Crab pulsar, the folded light curve (FLC) is obtained by 
estimating the phase (within the period) of each X-ray photon and forming a histogram of
counts as a function of phase. For such sources $\Delta T_d$ within a phase bin depends
not only upon the expected count rate in that bin, but also that in several earlier phase 
bins, depending upon the relative values of the detector deadtime and the width of 
the phase bin.  This can distort the observed FLC, which can be corrected by estimating 
$\Delta T_d$ as a function of phase, also known as phase resolved deadtime.

\begin{figure}
\centering
\includegraphics[width=9.2cm]{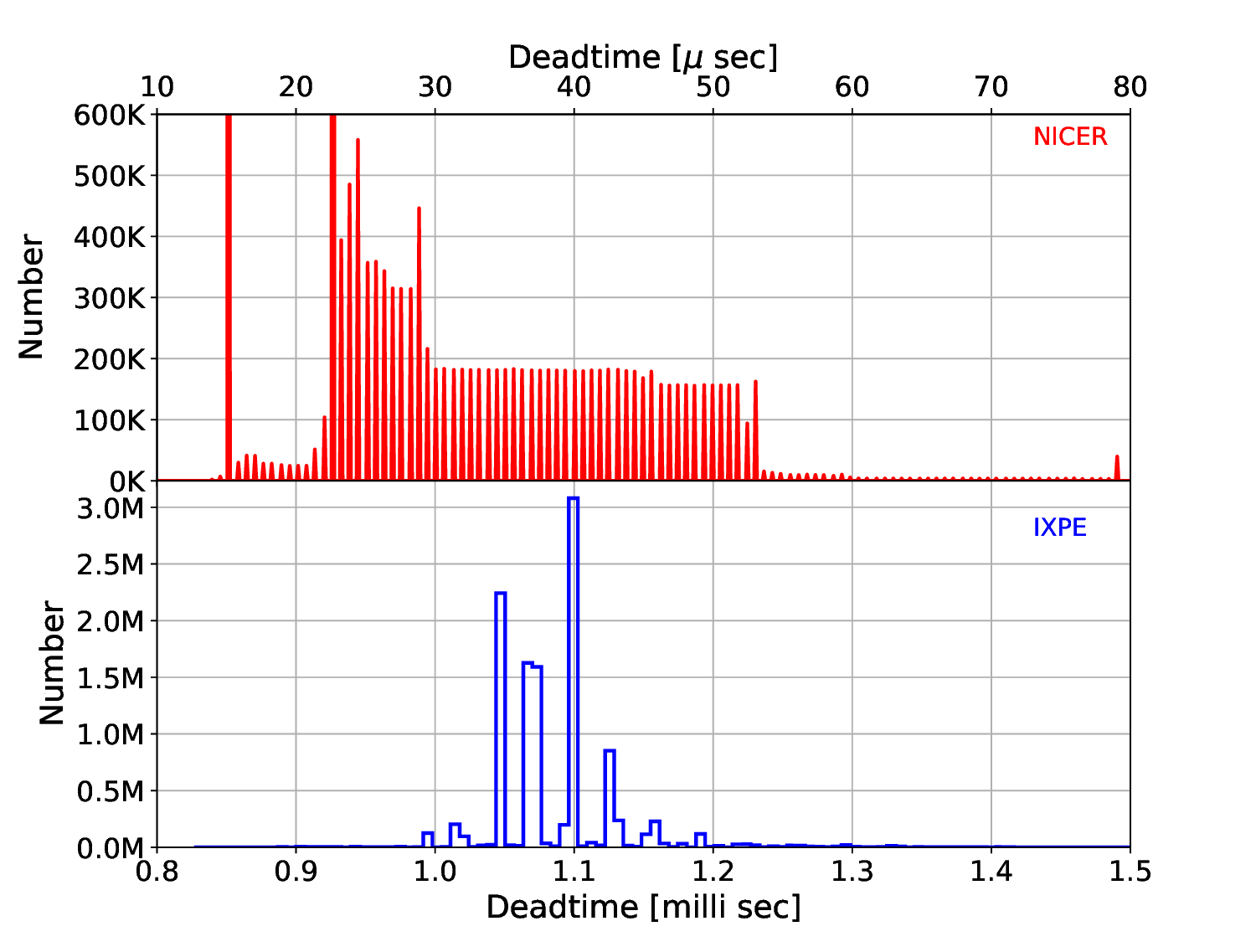}
\caption{
        Top panel: Histogram of deadtimes for a typical detector of the {{\it NICER}} X-ray 
        observatory for ObsID $1013010147$. Abscissa is in micro seconds ($\mu$ sec), the 
        ordinate is in units of thousand counts, and the bin width is $0.1$ $\mu$ sec. The 
        bins at abscissa $22.65$  and $15.15$ $\mu$ sec have $176,791,344$ and $14,306,804$ 
        counts respectively. Bottom panel: Histogram of deadtimes for detector DU1 of the 
        ObsID $02001099$ of the {{\it IXPE}} X-ray observatory. Abscissa is in milli seconds 
        (msec), the ordinate is in units of million counts, and the bin width is $7.2$ $\mu$
        sec.
        }
\label{fig1}
\end{figure}

Consider the deadtime of the the Neutron Star Interior Composition Explorer ({{\it NICER}}) 
satellite observatory \citep{Gendreau2017}. {{\it NICER}}'s design makes its deadtime
extremely low; the top panel of Fig.~\ref{fig1} shows the histogram of the deadtimes of a
typical detector of {{\it NICER}}. About $87.34$\% and $7.07$\% of the deadtimes lie around 
the values $22.65$  and $15.15$ $\mu$ sec, respectively \citep{LaMarr2016, Stevens2018, 
Vivekanand2020}; these two have not been shown in the figure, which shows the distribution of 
the rest. This results in a mean detector deadtime of $23.1 \pm 4.2$ $\mu$ sec 
\citep{Vivekanand2020}. For an average count rate of $211.3$ counts per sec per detector, 
the mean time interval between photons is about $4.7$ msec \citep{Vivekanand2020}. Clearly 
this is a case of the detector deadtime being much smaller than the mean time 
interval between X-ray photons from the source; Fig.~$8$ of \cite{Vivekanand2021} shows 
that the deadtime fraction is almost a replica of the expected light curve, and of a very 
small value, so the distortion of FLC is negligible.

Next consider The Nuclear Spectroscopic Telescope Array ({{\it NuSTAR}}) X-ray observatory
\citep{Harrison2013}. It has two detectors (Focal Plane Detector Modules) each of which 
has a deadtime of $2.5$ msec on account of event readout \citep{Harrison2013, Madsen2015}.
The average count rate per module for the Crab pulsar is $250$ counts /sec \citep{Madsen2015}
so the mean time between photons is $4$ msec. Now the detector deadtime is relatively
more comparable with the mean time interval between photons than in the {{\it NICER}} case. 
Further it is comparable to the width of the first peak of the FLC of the Crab pulsar. 
Therefore significant distortion of the FLC is expected; indeed the observed FLC in the 
top panel of Fig.~$3$ of \cite{Madsen2015} is a very distorted version of the corrected 
(expected) FLC shown in the second panel of the same figure.

Finally, consider the deadtime of one of the three detectors of the Imaging X-ray Polarimeter
Explorer ({{\it IXPE}}) \citep{Weisskopf2022}; the bottom panel of Fig.~\ref{fig1} shows the 
histogram of the deadtimes of detector DU1 of {{\it IXPE}}. The mean deadtime is $1.088 \pm 0.071$ 
msec\footnote{\url{https://heasarc.gsfc.nasa.gov/docs/ixpe/analysis/IXPE-SOC-DOC-008A\_UG-Instrument.pdf}}. 
The mean countrate for this observation is $\approx 75.3$ counts per sec over its entire energy band, 
implying a mean time interval between X-ray photons of $\approx 13.28$ msec. So this case is 
intermediate between the above two cases.

This work focuses on obtaining the phase resolved deadtime correction to the Crab pulsar's FLC
obtained from {{\it IXPE}} data. First, a simple analytical formula is derived that connects 
the observed and expected FLCs. This formula is verified using the first two cases. Then a 
method is given for obtaining the phase resolved deadtime correction for the IXPE data of the 
Crab pulsar, and then the results are verified as above. Finally, the importance of the phase 
resolved deadtime correction is discussed in connection with the X-ray emission mechanism of 
the Crab pulsar.

\section{Analytic formula for phase resolved deadtime} \label{sec:formula}

\begin{figure}
\centering
\includegraphics[width=9.2cm]{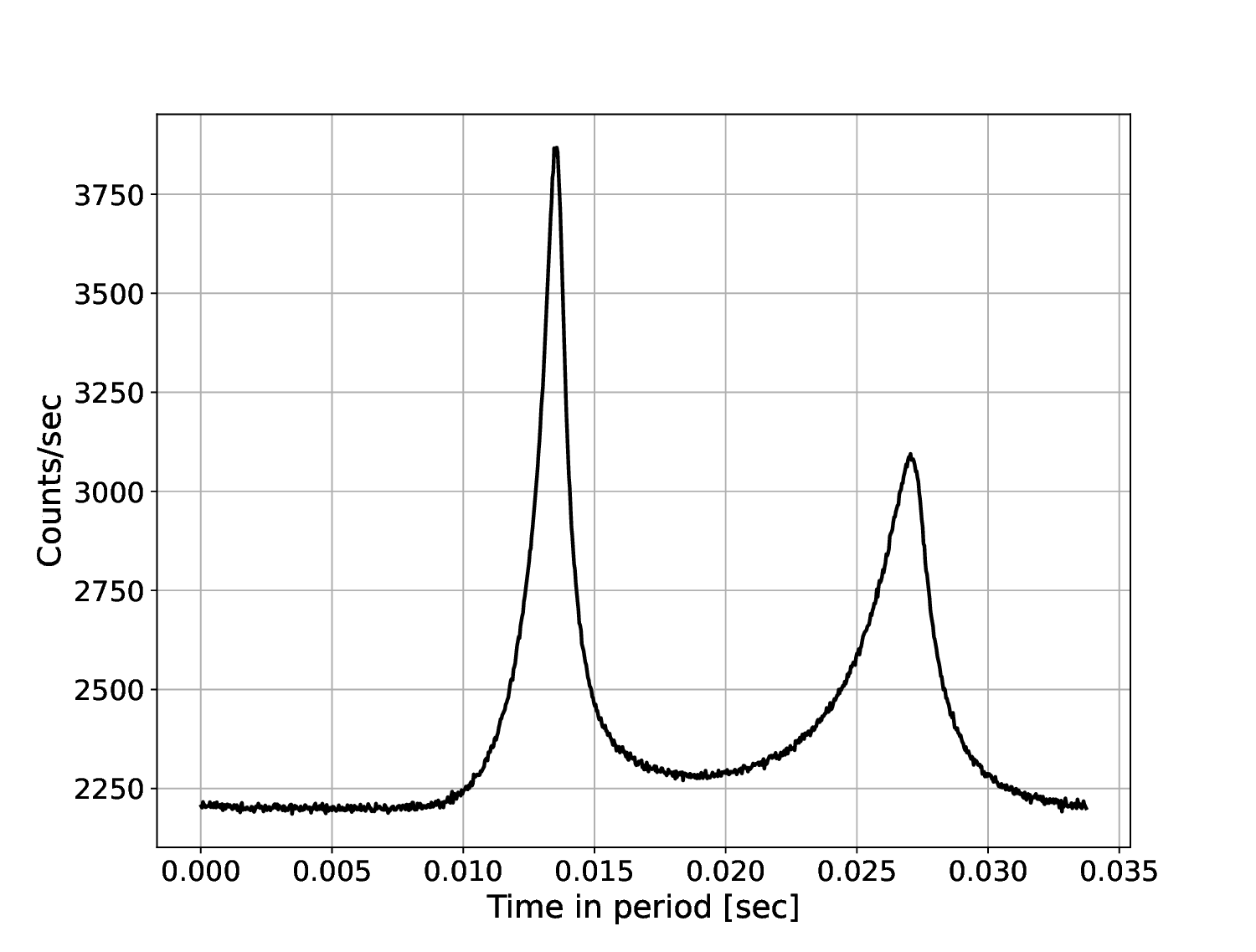}
\caption{
        Expected (i.e., deadtime corrected) FLC of the Crab pulsar using {{\it NICER}} 
        data of ObsIDs $1013010125$, $1013010126$, $1013010146$, $1013010147$, $1013010148$, 
        and $1013010150$, in the energy range $2 - 8$ keV. The abscissa has $1024$ time bins
        in the average period $P$ of these data ($P = 1/29.626728$ sec).
        }
\label{fig2}
\end{figure}

{{ Let $l_e(t)$ be the expected FLC of a pulsar which is a function of the time $t$ within 
the pulsar period $P$; over a duration of observation longer than $P$, $t \equiv t 
\ \text{modulo}\ P$. The time $t$ ranging from $0$ to $P$ is equivalent to the phase of the
periodic signal ranging from $0$ to $360$ degrees.}}
Let $l_o(t)$ be the observed FLC. For reference the 
$2 - 8$ keV $l_e(t)$ of the Crab pulsar from the {{\it NICER}} data of the six largest 
data files will be used, which is shown in Fig.~\ref{fig2}; details of its estimation
are given in \cite{Vivekanand2020}. 

Now consider a small time interval $\Delta t$ after a given time $t$. In an observation of 
total duration $T_0$ sec, there are $T_0 / P$ periods, so the expected and observed 
number of photons in the time interval $\Delta t$ are $(l_e(t) \times \Delta t) \times 
(T_0 / P)$ and $(l_o(t) \times \Delta t) \times (T_0 / P)$, respectively. Let $f_d(t)$ 
be the deadtime fraction per unit time at the time $t$. Then by definition,
\begin{equation} \label{eq1}
l_o(t) \Delta t (T_0 / P) = l_e(t) \left [ \Delta t \times ( 1 - f_d(t) ) \right ] (T_0 / P),
\end{equation}
\noindent
which implies that 
\begin{eqnarray} \label{eq2}
l_e(t) & = &  l_o(t) / (1 - f_d(t)) \nonumber \\
\Rightarrow f_d(t) & = & 1 - l_o(t) / l_e(t).
\end{eqnarray}
\noindent
Thus one can obtain the expected FLC from the observed FLC and the deadtime fraction, 
both of which are observed quantities.

For further derivation the following assumptions are made: (1) the detector deadtime 
$\tau$ is a constant; (2) $\tau \ll P$; (3) $l_e(t)$ and $l_o(t)$ are well defined 
and $> 0$; (4) $T_{ph}$ is the mean time interval between two subsequent photons; the 
observed and expected $T_{ph}$ are the inverse of the corresponding FLCs ($1 / l_o(t)$ 
and $1 / l_e(t)$).

\begin{figure}
\centering
\includegraphics[width=9.2cm]{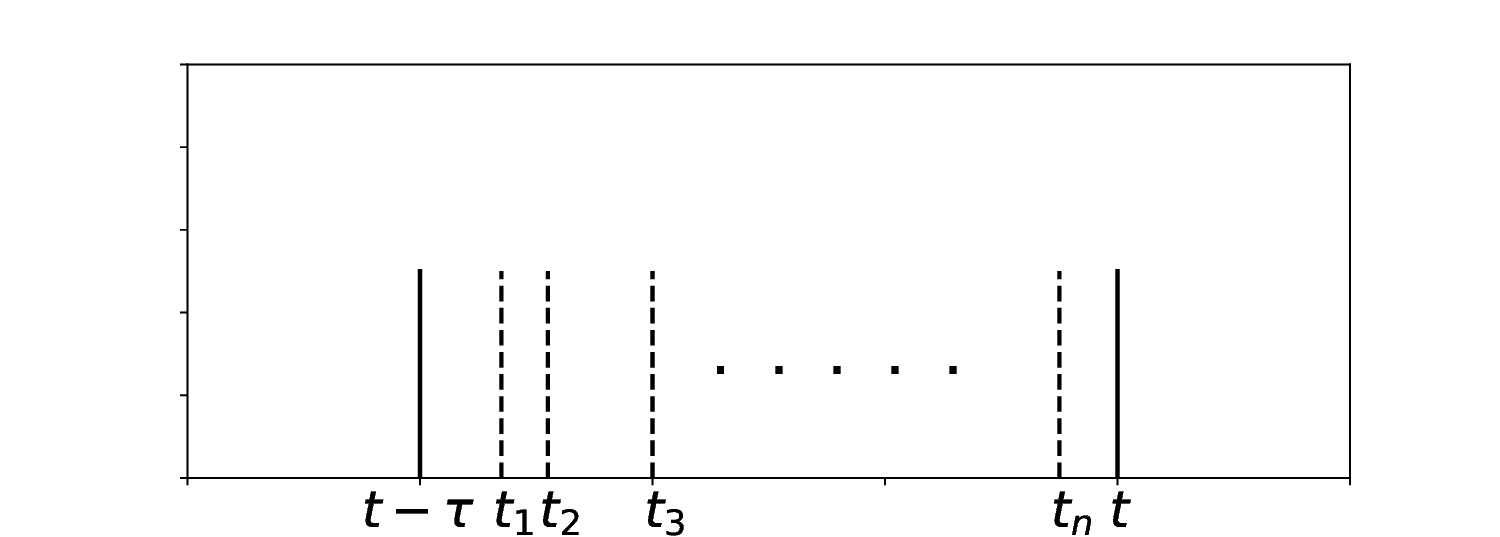}
\caption{
        Arrival times $t_1$, $t_2$, $t_3$, ..., $t_n$ and $t$ of $n + 1$ X-ray photons 
        in the duration $t - \tau$ to $t$.  These are determined by the expected FLC 
        $l_e(t)$ and Poisson statistics. The photon at time $t$ can be observed only 
        if the $n$ earlier photons were not received.
        }
\label{fig3}
\end{figure}

Let $P_e(t)$ be the expected normalized probability density of photon arrival at 
time $t$; the actual probability of its arrival in a small time interval 
$\epsilon$ is $P_e(t) \times \epsilon$. Clearly $P_e(t)$ depends upon $l_e(t)$. 
Now, the observed normalized probability density $P_o(t)$ depends not only upon 
$P_e(t) \epsilon$ but also the probability that no photon arrives in the time 
duration $t - \tau$ to $t$. This can be estimated by dividing this time duration 
$\tau$ into $M$ sub intervals of duration $\epsilon$, and taking the limits 
$\epsilon \rightarrow 0$ and $M = \tau / \epsilon$. Figure \ref{fig3} is a 
cartoon of the scenario being discussed here. Therefore the observed probability 
of receiving  a photon in the duration $\epsilon$ at time $t$ is given by
\begin{eqnarray} \label{eq3}
P_o(t) \epsilon & = & P_e(t) \epsilon  \times \left [ 1 - P_e(t - \epsilon) \epsilon \right ]  \nonumber 
                                       \times \left  [ 1 - P_e(t - 2 \epsilon) \epsilon \right ] \nonumber \\
                                  &  & \times \left  [ 1 - P_e(t - 3 \epsilon) \epsilon \right ] \nonumber 
                                        \dots \nonumber 
                                  \times \left  [ 1 - P_e(t - M \epsilon) \epsilon \right ] \nonumber \\
                & = & P_e(t) \epsilon \prod_{i = 1}^{M}  \left  [ 1 - P_e(t - i \epsilon) \epsilon \right ] \nonumber \\
                & \approx & P_e(t) \epsilon \left  [ 1 - \sum_{i = 1}^{M} P_e(t - i \epsilon) \epsilon \right ] \nonumber \\
                & \approx & P_e(t) \epsilon \left  [ 1 - \int_{t - \tau}^{t} P_e(t) dt \right ].
\end{eqnarray}
\noindent
In eqn~\ref{eq3} only terms linear in $\epsilon$ have been retained, and the summation has been 
converted into an integral.

Both $P_e(t)$ and $P_o(t)$ are defined above as being normalized. The common implication is that
$\int_{0}^{P} P_e(t) dt = 1$. However in our context $\int_{0}^{P} P_e(t)$ is equal to the average 
number of photons per period which is $\int_{0}^{P} l_e(t) dt$.

Now a model of $P_e(t)$ is required to connect it to $l_e(t)$. We will make the simple 
assumption that the actual probability $P_e(t) \epsilon  \approx \epsilon / (\alpha 
\times T_{ph})$ where $\alpha$ is a constant to be determined but which is expected to
be close to $\approx 1$. This assumption makes logical sense, it is 
linearly proportional to $\epsilon$, and finally it helps in deriving the simple analytic 
formula below. But $T_{ph} = 1 / l_e(t)$, so $P_e(t) \approx l_e(t) / \alpha$; i.e., the 
expected probability density $P_e(t)$ is linearly proportional to the expected FLC $l_e(t)$.
The same argument can be extended to the observed probability density $P_o(t)$, so that
\begin{eqnarray} \label{eq4}
l_o(t) & = & l_e(t) \times \left  [ 1 - \int_{t - \tau}^{t} ( l_e(t) / \alpha ) dt \right ] \nonumber \\
\Rightarrow f_d(t) & = & \int_{t - \tau}^{t} ( l_e(t) / \alpha ) dt.
\end{eqnarray}
\noindent
This is the simple formula that relates the observed and expected FLC at a given time to the
phase resolved deadtime fraction.

One can make the model more sophisticated by using the known distribution of $\tau$ 
instead of using a constant $\tau$. One can derive a more rigorous relation between 
$P_e(t)$ and $l_e(t)$ that may be non-linear, but that may not lead to the simple 
analytic formula of eqn~\ref{eq4}. As shown later, this simple formula is sufficient 
for our purpose.  The parameter $\alpha \approx 1$ for the {{\it NICER}} and {{\it IXPE}} 
instruments, and needs to be set to $\approx 2$ for {{\it NuSTAR}}, which has a 
significant distortion of $l_e(t)$ due to $\tau$.

An interesting aspect of eq.~\ref{eq4} is that for peaks in $l_e(t)$, the corresponding 
peaks in $f_d(t)$ are shifted towards larger times by the about $\approx \tau / 2$.
This can be understood by recognizing that the integral in eq.~\ref{eq4} is essentially 
a moving average filter. Now a conventional moving average filter would have an integral
of the kind $\int_{t - \tau/2}^{t + \tau / 2} ( l_e(t) / \alpha ) dt$, the limits of
integration being from $t - \tau/2$ to $t + \tau / 2$, which is symmetric around $t$.
However in eq.~\ref{eq4} the limits are asymmetric around $t$. It is easy to show 
therefore that if $l_e(t)$ was a symmetric pulse, like a Gaussian, then eq,~\ref{eq4} 
would give a symmetric pulse shaped deadtime fraction $f_d(t)$ whose peak is shifted by 
about $\approx \tau / 2$ for at least small $\tau$.

Equation~\ref{eq4} fails when the integral evaluates to $> 1.0$ which makes the deadtime 
fraction $f_d(t) > 1$, and the observed FLC $l_o(t)$ negative, both of which are 
unphysical. This occurs when either the deadtime $\tau$ is large, or the expected FLC 
$l_e(t)$ is large, or both. 

\subsection{Verification of formula using {{\it NICER}} data} \label{sec:formula-nicer}

For very small values of $\tau$, eqn~\ref{eq4} reduces to
\begin{eqnarray} \label{eq6}
f_d(t) & \approx & \tau l_e(t) / \alpha \nonumber \\
       & \approx & \tau l_e(t),
\end{eqnarray}
\noindent
if $\alpha$ is assumed to be $1$. Thus the deadtime fraction is linearly 
proportional to the expected FLC. This is clearly seen in Fig.~$8$ of 
\cite{Vivekanand2021}. The count rate at the first peak in the top panel of that 
figure is $318$, so eqn~\ref{eq4} gives $f_d(t) = 23.1 \times 318 / 1000000 = 
0.73$\%. The corresponding value in the bottom panel of that figure is 
$0.77$\%. The difference of $\approx 0.04$\% is due to the fact that the top 
panel of Fig.~$8$ of \cite{Vivekanand2021} consists of X-ray photons only
while the bottom panel has about $\approx 8$\% more counts that are not from 
the Crab pulsar but still contribute to the deadtime \citep{Vivekanand2021}.

Similarly the  count rate at the first sample in the top panel of that figure 
is $203$, so $f_d(t) = 23.1 \times 203 / 1000000 = 0.47$\%. The corresponding 
value in the bottom panel of that figure is $0.50$\%, the difference once again
being consistent with the additional $\approx 8$\% more counts contributing 
to the deadtime but not to the FLC.

In the {{\it NICER}} case the shift in the peak of $f_d(t)$ is not noticeable
because $\tau$ is smaller than the bin width. 

Thus the formula in eqn~\ref{eq4} has been verified for {{\it NICER}} data of
the Crab pulsar.

\subsection{Verification of formula using {{\it NuSTAR}} data} \label{sec:formula-nustar}

The ideal starting point for verifying eqn~\ref{eq4} is the actual expected FLC $l_e(t)$
which is obviously not available; only the observed FLC $l_o(t)$ is available.
However for {{\it NICER}} data of the Crab pulsar the deadtime $\tau$ is very 
small, so the difference between $l_e(t)$ and $l_o(t)$  is also very small.
So in the previous sub-section we used the deadtime corrected $l_o(t)$ as 
the actual $l_e(t)$ for {{\it NICER}} data, shown in the top panel of Fig.~$8$ 
of \cite{Vivekanand2021}, which includes the entire energy range of the 
{{\it NICER}} observatory. 

For this section also we ideally need to use the actual $l_e(t)$ which is
not available. Since $\tau$ for {{\it NuSTAR}} data is quite large, significant
differences are expected between $l_o(t)$ and the actual $l_e(t)$. So we can
not use the estimated $l_e(t)$ given in the middle panel of Fig.~$3$ of 
\citep{Madsen2015}, to reproduce the $l_o(t)$ in the top panel of that figure, 
using eqn~\ref{eq4} above. Therefore we can not do any quantitative comparison
here. Therefore in this section we will use the $l_e(t)$ in Fig.~\ref{fig2} 
above since only a qualitative comparison is attempted here. This figure was 
prepared for verification of the {{\it IXPE}} case which follows later. The
use of Fig.~\ref{fig2} in this section is discussed in greater detail later
on.

The mean X-ray flux of the Crab pulsar per detector of the {{\it NuSTAR}} 
observatory is $\approx 250$ counts/sec \citep{Madsen2015}, while at the 
first peak it is $\approx 400$ counts/sec. The corresponding observed 
$T_{ph}$ are $4$ and $2.5$ msec respectively. This is a case of the detector
deadtime ($2.5$ msec) being comparable to the $T_{ph}$. This reduces the 
accuracy of the formula in eqn~\ref{eq4}. However the formula 
of eqn~\ref{eq4} does quite well with the {{\it NuSTAR}} data if $\alpha$ is 
chosen to be $\approx 2$ instead of $1$; its implications will be discussed 
later on.

\begin{figure}
\centering
\includegraphics[width=9.2cm]{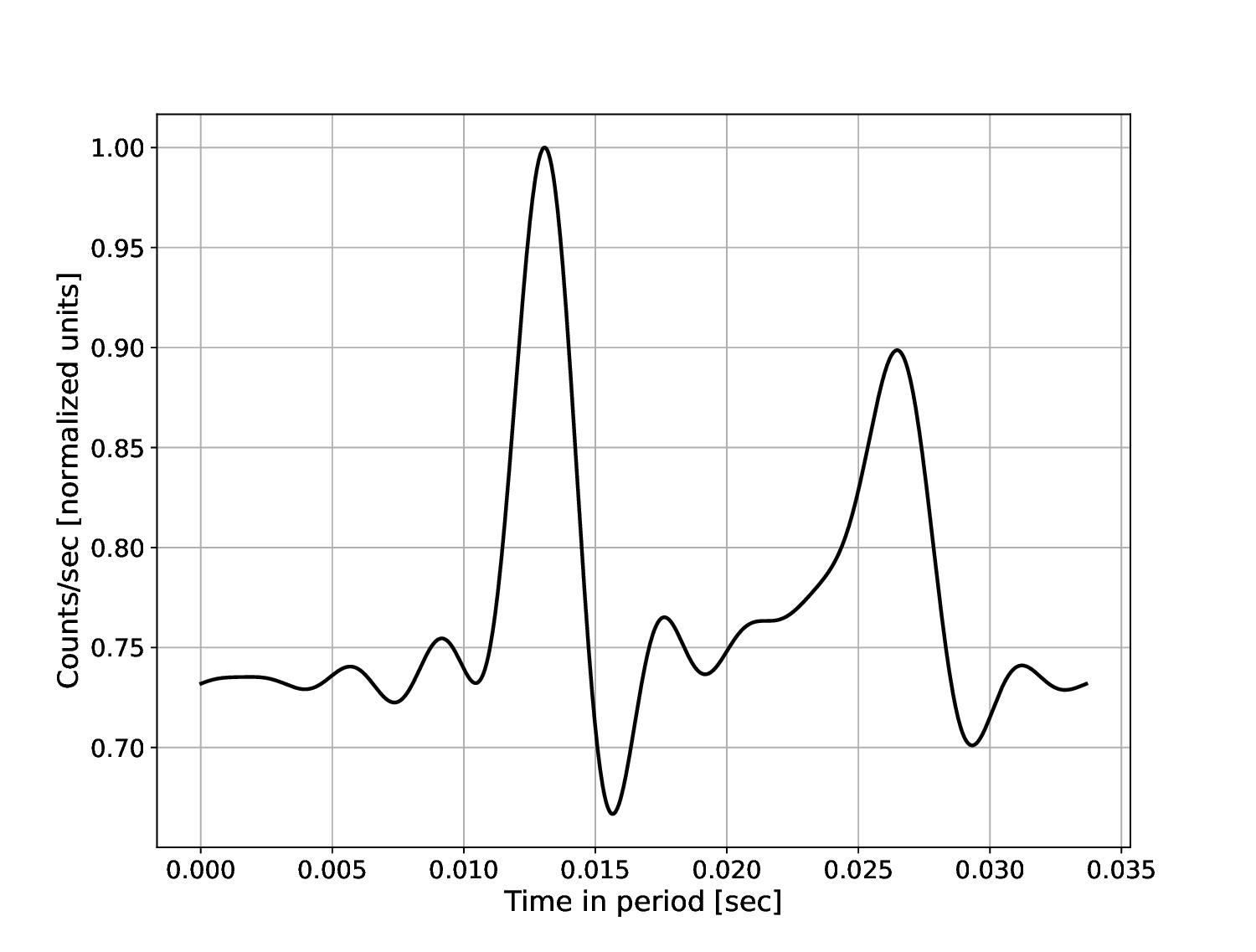}
\caption{
        The FLC obtained by using eqn~\ref{eq4} on a scaled version of
        Fig.~\ref{fig2} and using a low pass filter of smoothing time
        $3.4$ msec.
        }
\label{fig4}
\end{figure}

We start by assuming that the FLC in Fig.~\ref{fig2} is the actual $l_e(t)$ for the
{{\it NuSTAR}} data of the Crab pulsar if it is scaled to reflect the mean 
counts per sec. This data is smoothed by implementing a moving average filter
over $5$ samples to approximately reflect the number of samples per period in 
the top panel of Fig.~$3$ of \cite{Madsen2015}. Then the data is scaled to 
give an average count rate of $250$ counts/sec. Then eqn~\ref{eq4} is used 
with $\tau = 2.5$ msec and $\alpha = 2$ to obtain the observed FLC $l_o(t)$. 
This data is then passed through a box shaped low pass filter of time constant 
$3.4$ msec to produce the damped oscillations observed by \cite{Madsen2015} in 
the top panel of their Fig.~$3$. The result is shown in Fig.~\ref{fig4}, which 
compares very well with the corresponding curve in the top panel of Fig.~$3$ 
of \cite{Madsen2015}. 

With the peak in Fig.~\ref{fig4} normalized to the value $1.0$, the first and 
second minima at $t = 0.01567$ and $t = 0.02932$ are $0.67$ and $0.70$ 
respectively; these are consistent with the values in the top panel of Fig.~$3$ 
of \cite{Madsen2015}. The secondary maximum at $t = 0.02649$ is $0.90$, which 
is also consistent with the corresponding value in top panel of Fig.~$3$ of 
\cite{Madsen2015}. The only difference is the damped oscillation before the 
first peak, which is missing in their Fig.~$3$; this difference is discussed in 
greater detail later on. {{ Further, \cite{Madsen2015} note in the legend of 
their Fig. $3$ that ``the minimum of the live-time curve occurs just after the 
peak", as expected from eqn~\ref{eq4}, by an amount that they do not specify, 
but which appears to be at least $0.7$ ms by visual analysis. While this is not
entirely consistent with the expected value of $2.5 / 2 = 1.25$ ms, it is in the 
correct ball park.}}

Thus Fig.~\ref{fig4} is a reasonable verification of eqn~\ref{eq4} for the Crab 
pulsar data of {{\it NuSTAR}} (assuming that $\alpha = 2$), even though the 
deadtime $\tau$ in this case is relatively large.

Figure~\ref{fig2} contains data in the energy range $2$ to $8$ keV while the energy
range of {{\it NuSTAR}} is from $3$ to $78$ keV \citep{Madsen2015}. However the Crab
pulsar has a very steep power law spectrum \citep{Vivekanand2021} so most of the
photons in the FLC are at the lower energies. So we can ignore the spectral evolution 
of the Crab pulsar's FLC for our qualitative analysis.

In section~\ref{sec:formula} we made the assumption that $P_e(t) \epsilon  \approx 
\epsilon / (\alpha \times T_{ph}) \approx \epsilon l_e(t) / \alpha$. Since $\tau$ is
large in the {{\it NuSTAR}} case, the count rate of the actual light curve needs to be 
lowered by setting $\alpha = 2$ to get sensible results. This can by no means be 
considered a rigorous ``fix'' of the formula, but Fig.~\ref{fig4} shows that this ``fix''
achieves quite accurate qualitative results. The ideal solution is to get a more
accurate relation between $P_e(t)$ and $l_e(t)$.

\section{Deadtime estimation for {{\it IXPE}}} \label{sec:ddtixpe}

A very brief introduction to {{\it IXPE}} is given in its quick-start guide\footnote{\url{https://heasarc.gsfc.nasa.gov/docs/ixpe/analysis/ixpe\_quickstart\_v2.pdf}}. More details are provided in its observatory\footnote{\url{https://heasarc.gsfc.nasa.gov/docs/ixpe/analysis/IXPE-SOC-DOC-011A\_UG-Observatory.pdf}} and instrument\footnote{\url{https://heasarc.gsfc.nasa.gov/docs/ixpe/analysis/IXPE-SOC-DOC-008A\_UG-Instrument.pdf}} guides. 

For our purpose it is sufficient to know that there are three gas pixel detectors (GPD) each of which 
produces a track after absorbing an X-ray photon; this track contains information about the polarization
of that photon. The deadtime of the detector depends upon the time required to read the electrons in
this track; section $2.2$ of the instrument guide discusses this in detail. An example of deadtime
distribution is given in the bottom panel of Fig.~\ref{fig1}. The deadtime is recorded in the level-1
data files of {{\it IXPE}} in terms of the parameter LIVETIME, which is the detector live time since the 
previous event. Let two successive events occur at the {{\it IXPE}} detector at times t1 and t2, with 
LIVETIMEs l1 and l2, respectively. Then the deadtime after the first event is (t2 -t1) - l2, since the 
interval between any two successive events is the sum of the deadtime after the first event and the live 
time until the next event.

However the events recorded in the level-2 files are often not successive, since this data has gone through
several filters such as passage of the observatory through the South Atlantic Anomaly (SAA), entering Earth 
occultation, etc. Thus, the event-by-event LIVETIME becomes useless and can not be carried forward to level-2
files. So eqn.~\ref{eq4} is applicable to data in the level-1 files only, which include all events,
since ``Level-1 Science Event data contain the same information as Level-0 Science Event data but converted 
to FITS format"\footnote{\url{https://heasarc.gsfc.nasa.gov/docs/ixpe/analysis/IXPE-SOC-DOC-007D\_UG-DataFormats.pdf}}.
Therefore in this section the {{\it IXPE}} level-2 data files are used for obtaining the observed FLC while
the corresponding level-1 data files are used to derive the phase resolved deadtime fraction. Note that the
level-1 files {{ probably}} can not {{ be}} used to obtain the observed FLC since these contain events that are 
not related to the X-ray source but nevertheless contribute to the detector deadtime. Further, both level-1 
and level-2 data files have to be referred to the barycenter frame for phase resolved analysis.

Currently data of three ObsIDs are available for {{\it IXPE}} -- $01001099$, $02001099$ and $02006001$.
Data of ObsID $01001099$ has problems described by \cite{Bucciantini2023} with details in the README file 
associated with this paper. ObsID $02006001$ has $\approx 2.5$ times less data than ObsID $02001099$,
so only the latter has been considered for analysis in this section. The average count rate per {{\it \
IXPE}} detector for the Crab pulsar is $\approx 44$ counts/sec in the energy range $2$ to $8$ keV, but 
differs slightly from detector to detector.

\begin{table}
\begin{center}
\caption{ Crab pulsar's rotation parameters for obtaining the observed FLC
from the data of OBS\_ID $02001099$. \label{tbl1}
}
\begin{tabular}{|l|c|}
\hline
Parameter  & Value \\
\hline
Epoch (MJD) & $60017.226732$ \\ 
\hline
$\nu_0$ (Hz) & $29.5746291403(9)$ \\
\hline
$\dot \nu_0$ ($10^{-10}$ Hz s$^{-1}$) & $-3.67074(1)$ \\
\hline
$\ddot \nu_0$ ($10^{-21}$ Hz s$^{-2}$) & $8.5(2)$ \\
\hline
\hline
\end{tabular}
\end{center}
\end{table}

The algorithm for the analysis of this section is the following:
\begin{enumerate}
\item There is a level-2 data file for each of the three detectors. Barycenter these epochs, combine them
and obtain the best fit rotation frequency $\nu_0$ of the Crab pulsar and its first and second derivatives 
$\dot \nu_0$ and $\ddot \nu_0$ using the stride-fit method \citep{Vivekanand2020}. The second derivative
is required since this data duration is more than $40$ days. The results are shown in Table~\ref{tbl1}. 
Further analysis has to be done separately for each detector.
\item For each detector there are two level-1 files. For each of them, there are two Good Time Interval (GTI) 
extensions. These two GTIs are merged with the GTI of the corresponding level-2 data file using the
AND mode, i.e., output GTI contains time intervals common to the three input GTI (also known as the 
intersection set). Note that the second GTI extension has an extra column named RUN\_ID which should be 
deleted beforehand.
\item Then the level-1 data files are filtered using the merged GTI extension. Sequential pairs of the
TIME and LIVETIME values are processed to yield three epochs, viz., the epoch of the current event, 
this epoch plus the deadtime, and the epoch of the next event, the condition being that the pair must 
belong to the same GTI.
\item Finally these three epochs are referred to the barycenter.
\end{enumerate}

\begin{figure}
\centering
\includegraphics[width=9.2cm]{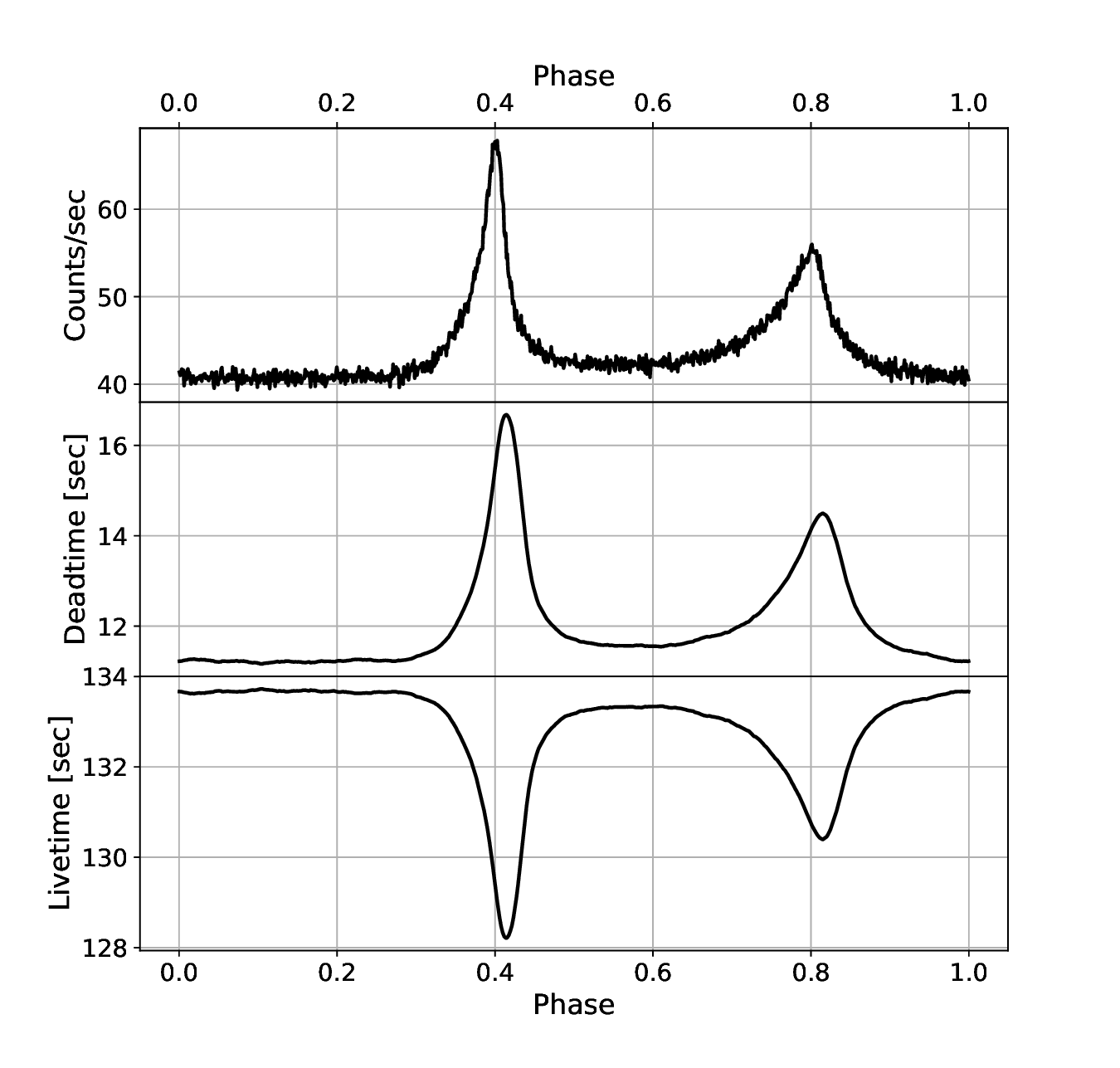}
\caption{
        Top panel: Observed FLC of the Crab pulsar in the energy range $2 - 8$ keV
        from Detector $1$ of ObsID $02001099$ of {{\it IXPE}} level-2 data. Middle 
        panel: Phase resolved deadtime of the same detector from level-1 data. 
        Bottom panel: Phase resolved livetime of the same detector from level-1 data.
        }
\label{fig5}
\end{figure}

The result of the above algorithm is that one has two output files for each of the three detectors,
each file containing three columns of barycentered times -- the epoch of an event, the epoch of the next 
sequential event, and an intervening epoch signifying the end of the deadtime after the first event, or 
equivalently the start of the livetime before the next sequential event. Clearly the difference between 
the times of the intervening and first events is the deadtime at that epoch, while the difference between 
the times of the third and intervening events is the corresponding livetime. By estimating the phase 
within the Crab pulsar period of the first event, one can make a histogram of the deadtime and livetime 
as a function of phase within the period; this has been done for {{\it NuSTAR}} by \cite{Madsen2015} and 
for {{\it NICER}} by \cite{Vivekanand2020, Vivekanand2021}.

Figure~\ref{fig5} shows the result of applying this algorithm on the data of detector
DU1 of ObsID $02001099$ of {{\it IXPE}} data. The top panel shows the observed FLC of
the Crab pulsar using the level-2 data, while the middle and bottom panels show the
phase resolved deadtime and livetime, respectively, estimated from the level-1 data.
The peak of the deadtime is shifted by about $1.088 / 1000 / 2 * 29.626728 \approx 
0.016$ in phase with respect to the peak of the FLC, as expected from eqn~\ref{eq4}.
The deadtime and livetime are perfectly anti correlated as expected, since their sum 
is the total time spent observing the source at that phase; this quantity is 
independent of phase, its average value per phase bin being $144.9 \pm 0.3$ sec.

The phase resolved deadtime fraction (observed) can be obtained from Fig.~\ref{fig5} 
by the formula $f_d(t) = $\ deadtime / (deadtime + livetime). {{ The minimum, 
maximum and mean values of the deadtime fraction are $7.70$\% (per cent), $11.51$\%
and $8.29$\% respectively; the error on these quantities is $\approx 0.02$\%.
\cite{Bucciantini2023} mention that the difference between the maximum 
and minimum deadtime fractions ``is estimated to be less than $3$\%"; the corresponding 
number here is $11.51 - 7.70 = 3.81$\%. Thus the result of \cite{Bucciantini2023} can be
considered to be preliminary but in the right ball park, and any further study requires 
a more rigorous calculation.}}

\begin{figure}
\centering
\includegraphics[width=9.2cm]{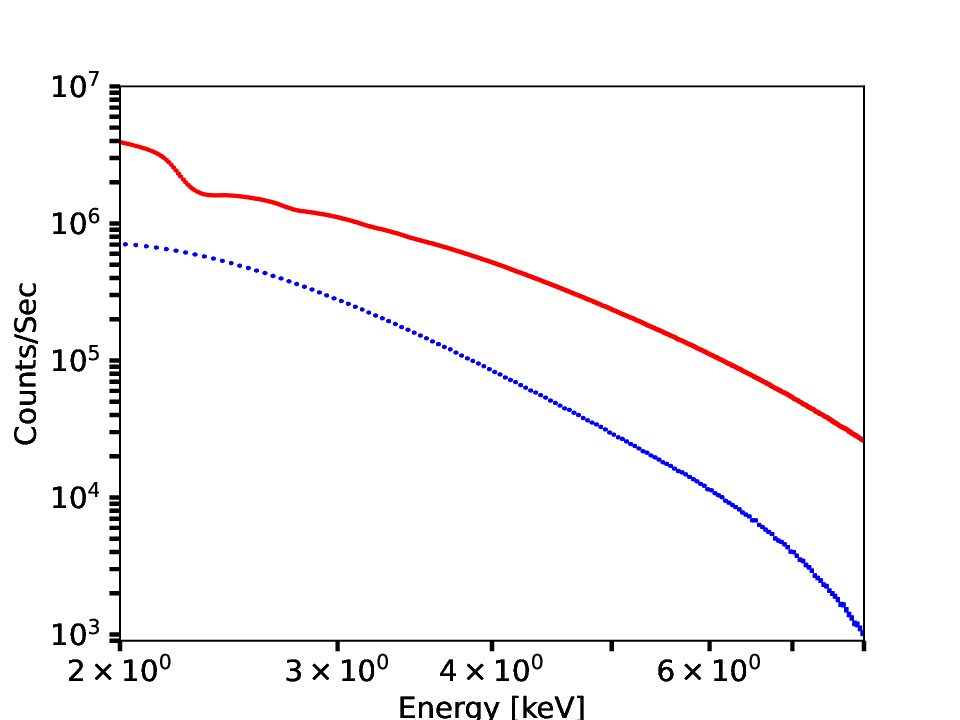}
\caption{
        Raw spectrum of the Crab pulsar in the energy range $2$ to $8$ keV
        using the {{\it NICER}} (red curve) and {{\it IXPE}} (blue curve)
        data.
        }
\label{fig6}
\end{figure}

Now we need to use eqn.~\ref{eq4} on the actual $l_e(t)$ of {{\it IXPE}} data to
estimate the expected $f_d(t)$. However this is not available. So we instead use
the FLC in Fig.~\ref{fig2} which pertains to the {{\it NICER}} data in the same
energy range as {{\it IXPE}}. Now the FLC of the crab pulsar evolves with energy.
Figure~\ref{fig6} shows the raw spectrum of the {{\it NICER}} and {{\it IXPE}} 
data in the same energy range; the spectra are roughly of the same shape in this
energy range, but some differences exist between them, which will be discussed
later on. So we will go ahead and use Fig.~\ref{fig2} as the actual $l_e(t)$ for 
this section.

\begin{figure}
\centering
\includegraphics[width=9.2cm]{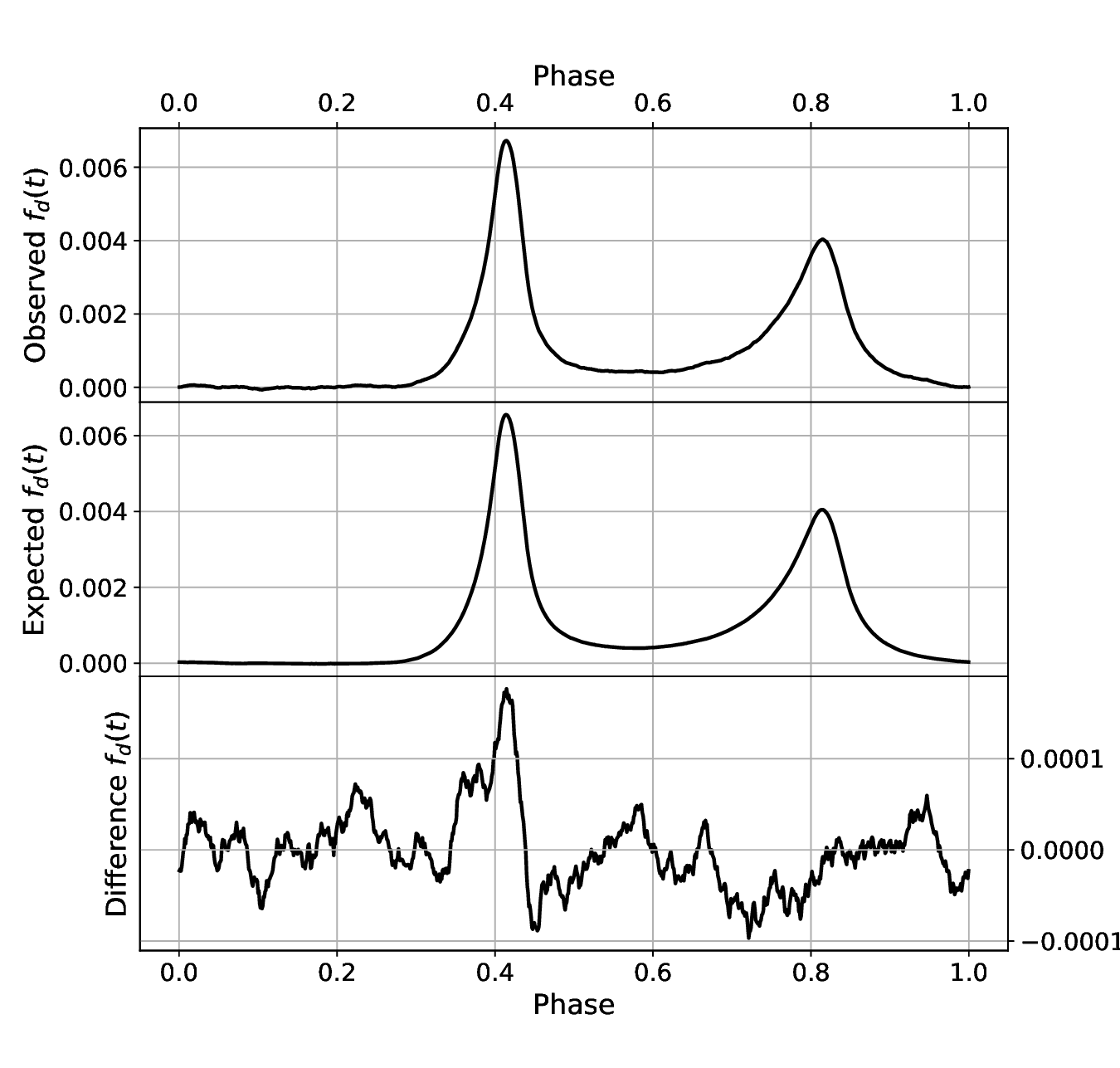}
\caption{
        Top panel: Observed $f_d(t)$ derived from the second and third panels of 
        Fig.~\ref{fig5}, after removing the off-pulse mean and normalizing the 
        area under the curve to $1.0$. Middle panel: Expected $f_d(t)$ estimated 
        using eqn.~\ref{eq4} on the data of Fig.~\ref{fig2}, after mean removal 
        and area normalization. Bottom panel: the difference between the top and
        middle panels.
        }
\label{fig7}
\end{figure}

From Fig.~\ref{fig5} the observed $f_d(t)$ should be compared with the expected 
$f_d(t)$ obtained by using eqn.~\ref{eq4} (with $\alpha = 1$) on the data of 
Fig.~\ref{fig2}, after scaling for the mean count rate of $\approx 44 / (1 - 
0.083) \approx 48$ photons/sec for a single detector. A direct comparison of 
these two curves proves difficult because the Crab nebula background counts are 
significantly different in the  {{\it NICER}} and {{\it IXPE}} cases. So the nebular 
background was estimated in both $f_d(t)$ curves by estimating the mean value of the 
counts in the off-pulse phases ($0.0$ to $0.2$). After subtraction of the nebular 
background the area under each $f_d(t)$ curve was normalized to the value $1.0$. These 
two curves are shown in the top two panels of Fig.~\ref{fig7}; the bottom panel shows 
their difference.

We identify three phase regions for comparison -- region $1$ being under the first peak,
extending from phase $0.35$ to $0.45$, region $2$ being under the second peak, extending 
from phase $0.75$ to $0.85$, and region $3$ being the rest of the phases. The mean and
standard deviation of the difference curve in region $3$ are $-5.7 \times 10^{-6}$ and 
$31.5 \times 10^{-6}$ respectively. Using these values the $\chi^2$ per degree of freedom
in the three regions are $10.6$, $1.1$ and $1.0$ respectively. The reduced $\chi^2$ is 
$1$ for region $3$ as expected, and is close to $1$ under the second peak, but significantly
higher than the expected value of $\approx 1$ under the first peak.

The difference under the first peak could be due to two factors -- lack of alignment of the
observed and expected $f_d(t)$ curves at the level of fraction of a phase bin, which would
results in a sharp rise and fall of the difference curve under the first peak; or energy
evolution of the FLC due to the slightly differing raw spectra in Fig.~\ref{fig6}. To test
the first hypothesis, one of the curves was shifted with respect to the other by various 
fractions of the bin width, using Fourier techniques, but no improvement was noticed.
This leaves the second possibility, that our choice of Fig.~\ref{fig2} for the actual
$l_e(t)$ of {{\it IXPE}} is not completely valid.

There might also be issues related to alignment of the curves because of insufficient
number of photons in the FLC of the {{\it IXPE}} data, which was aligned with the FLC
of {{\it NICER}} data in Fig.~\ref{fig2}. Attempt to align the two curves showed that
the mis-alignment is smaller than the bin width of the data. Thus the alignment can be
improved only by obtaining more {{\it IXPE}} data of longer observations on the Crab 
pulsar. In fact it is for this very reason that data of the third ObsID ($02006001$) 
could not be used, since its data was too meager for FLC alignment.

The qualitative nature of the difference $f_d(t)$ for the other two detectors (DU2 and DU3)
of {{\it IXPE}} is similar to that of detector DU1, the sharp swing under the peak being
prominent. For DU2 the $\chi^2$ per degree of freedom in the three regions are $13.3$ $2.2$ 
and $1.0$; for DU3 the values are $24.9$, $2.8$ and $1.0$.

While the $\chi^2$ give one kind of information, it can be noticed that the maximum
departure of the difference curve in region $1$ is $0.00017 / 0.00658 = 0.026$, which is not 
too different from the Poisson error due to photon counts of $\approx 0.01$ at the peak of the FLC. 
So the analysis of this section may be facing problems of both energy evolution of the FLC, 
and low photon counts.

Thus Fig.~\ref{fig7} is a reasonable verification of eqn~\ref{eq4} for the Crab pulsar data 
of {{\it IXPE}} (assuming that $\alpha = 1$).

\section{Discussion} \label{sec:disc}

Finally, the corrected FLC for each detector is obtained by using the observed deadtime fraction 
$f_d(t)$ and the observed FLC in eqn.~\ref{eq2}. This has to be done separately for each detector 
since there can be minor differences between the parameters of each detector. For example, the
average deadtimes for the three detectors are $1.088 \pm 0.071$ msec, $1.096 \pm 0.074$ msec, and 
$1.094 \pm 0.074$ msec, which are almost the same. However their mean count rates are $75.25 \pm 
0.02$, $70.66 \pm 0.02$ and $67.76 \pm 0.02$ respectively. their maximum difference is $(75.25 - 
67.76) / 75.25 \approx 10$\%, which is not insignificant in our context since the observed FLC 
$l_o(t)$ in eqn.~\ref{eq2} would be proportional to these values.

A critical assumption in the above analysis is that one is processing pairs of successive events.
However this is not guaranteed; some events may be missing or filtered out for various reasons.
Thus if one had three successive events at times t1, t2 and t3, with LIVETIMEs l1, l2 and l3 
respectively, and if the intervening event was missing, one would be processing events t1 and t3,
and the derived deadtime and livetime for this pair would be in error, depending upon what values
were recorded as LIVETIME. The maximum deadtimes estimated for each detector are $7.367$, $7.806$
and $7.421$ msec respectively; clearly these extreme values are not part of statistical distributions
whose mean value is $\approx 1.09$ msec with standard deviation of $\approx 0.07$ msec. Therefore
it is possible that some of these extreme values could be erroneous deadtime estimates. However
the number of these extreme deadtime values is insignificant, as seen in the lower panel of
Fig.~\ref{fig1}.

The average livetimes of the three detectors are $12.043$, $12.886$ and $13.480$ msec; their 
inverses are $83.04$, $77.60$ and $74.18$ counts/sec respectively. These differ from the
observed count rates above by the expected correction for the deadtime, as expected. The maximum 
livetime values estimated for the three detectors are $180.682$, $217.177$, and $231.013$ msec
respectively. Once again some of these extreme livetime estimates may be {{ fake}}, because
it is unlikely that {{ no photons are detected by the DU}} in a duration of $\approx 200$ msec 
when the mean time between photons {{ is expected to be}} $\approx 12$ to $13$ msec.

Attempt was made to suppress the damped oscillation before the first peak in Fig.~\ref{fig4}, 
to be consistent with the top panel of Fig.~$3$ of \cite{Madsen2015}. Various integration
times were used in eqn.~\ref{eq4} along with different filter time constants, but the damped 
oscillations behaved similarly at both peaks; it was not possible to suppress only one of 
them. So it is possible that some other physical effect is operative here.

The dependence of deadtime upon photon energy was tested by plotting LIVETIME against the 
level-1 equivalent of photon energy PHA and PHA\_EQ. Scatter plots were made between PHA and 
PHA\_EQ, and PHA and LIVETIME. The former pair of parameters were highly correlated for each 
detector as expected, since they are essentially the same quantity. The latter pair showed 
no correlation, even when the LIVETIME at an epoch was correlated with the PHA of the
epoch just before that. It is therefore concluded that there is no energy dependence of the 
deadtime. {{ However this could also arise due to the fact that spectrum is dominated by photons 
in a narrow energy range of $2 - 4$ keV, and also possible by the poor photon statistics.}}

\subsection{X-ray emission mechanism of the Crab pulsar} \label{sec:emission_mech}

The characteristic feature of the Crab pulsar's X-ray FLC (Fig.~\ref{fig2}) is its first peak,
also known as the main pulse. This luminous and narrow peak is believed to be due to the 
formation of caustics in the high energy emission region of the Crab pulsar. Caustics are 
formed when photons from different regions of the pulsar magnetosphere arrive at the same 
time at the observer, which is believed to be due to a combination of special relativistic 
aberration and light travel time (see \cite{Romani1995}, \cite{Bai2010} and \cite{Harding2016} 
and references therein).

\cite{Romani1995} reproduce the fast sweep of the position angle of linear polarization of 
the Crab pulsar in relation to its first peak in the FLC (see their Fig.~$5$), and it is 
consistent with the optical polarization data of the Crab pulsar of \citep{Smith1988} and
\cite{Słowikowska2009}. Both works show that the percentage of linear polarization of the 
Crab pulsar at optical wavelengths is not minimum under the first peak of optical FLC, but 
is situated a bit beyond it in phase. In fact \cite{Słowikowska2009} state that it is 
minimum at the peak of the radio FLC.

To study the X-ray emission mechanism of the Crab pulsar one needs both an X-ray FLC and
phase resolved X-ray polarization data with sufficient signal to noise ratio and high phase 
resolution ($>= 1024$ bins in the period); {{\it IXPE}} is capable of providing such data.
If caustics form in the Crab pulsar's FLC as postulated, then a sharp peak in the FLC is 
expected where photons from different regions of the magnetosphere arrive at the same time 
at the observer. By the same token, the percentage of linear polarization should decrease 
under the sharp peak, since photons from different regions of the magnetosphere will have 
differing position angles of linear polarization, depending upon the geometry of the magnetic 
field and the specific high energy emission mechanism. This difference, between the phase of 
the maximum of the first peak of the X-ray FLC and the phase of the minimum of the percentage 
of X-ray linear polarization, may be an important constraint for the X-ray emission mechanism 
of the Crab pulsar. Correcting the observed X-ray FLC using the observed deadtime fraction 
$f_d(t)$ is an important step in this direction.

This work has made use of the publicly available data from the {{\it NICER}} and {{\it IXPE}}
X-ray observatories.

\begin{acknowledgments}
I thank the {{\it IXPE}} helpdesk for discussion regarding the instrumental issues of this work.
{{ I thank Niccolo Bucciantini and the referee for helpful comments.}}
\end{acknowledgments}


\begin{thebibliography}{}
\bibitem[Bai \& Spitkovsky (2010)]{Bai2010} Bai, X.N., Spitkovsky, A., 2010, \apj, 715, 1270
\bibitem[Bucciantini et al (2023)]{Bucciantini2023} Bucciantini, N. Ferrazzoli, R., Bachetti, M. et al 2023, NatAs, 7, 602
\bibitem[Gendreau \& Arzoumanian (2018)]{Gendreau2017} Gendreau, K. \& Arzoumanian, Z. 2017, NatAs, 1, 895
\bibitem[Harding (2016)]{Harding2016} Harding, A. 2016, J. Plasma Phys. 82, 635820306
\bibitem[Harrison et al. (2013)]{Harrison2013} Harrison, F.A., Craig, W. W., Christensen, F. E.,
et al. 2013 \apj 770, 103
\bibitem[LaMarr et al (2016)]{LaMarr2016} LaMarr, B., Prigozhin, G., Remillard, R., et al 2016, Proc. SPIE, 9905, 99054W
\bibitem[Madsen et al. (2015)]{Madsen2015} Madsen, K. K., Reynolds, S., Harrison, F., et al. 2015, \apj, 801, 66
\bibitem[Romani \& Yadigaroglu (1995)]{Romani1995} Romani, R.W. \& Yadigaroglu, I.A. 1995, \apj, 438, 314
\bibitem[Słowikowska et al (2009)]{Słowikowska2009} Słowikowska, A., Kanbach, G., Kramer, M., et al 2009, \mnras, 397, 103
\bibitem[Smith et al (1988)]{Smith1988} Smith, F.G., Jones, D.H.P., Dick, J.S.B., et al 1988, \mnras, 233, 305
\bibitem[Stevens et al (2018)]{Stevens2018} Stevens, A.L., Uttley, P., Altamirano, D., et al 2018, \apjl, 865, L15
\bibitem[Vivekanand (2020)]{Vivekanand2020} Vivekanand, M. 2020, \aap, 633, A57
\bibitem[Vivekanand (2021)]{Vivekanand2021} Vivekanand, M. 2021, \aap, 649, A140
\bibitem[Weisskopf et al (2022)]{Weisskopf2022} Weisskopf, M. C., Soffitta, P., Baldini, L. et al 2022, 
Journal of Astronomical Telescopes, Instruments, and Systems, Vol. 8, No. 2, article id. 026002
\end{thebibliography}
\end{document}